\documentstyle[aps]{revtex}
\begin{document}
\newcommand{\beq}{\begin{equation}}
\newcommand{\eeq}{\end{equation}}
\newcommand{\bea}{\begin{eqnarray}}
\newcommand{\eea}{\end{eqnarray}}
\def\plumin{\underline{+}}
\def\minplu{{{\stackrel{\underline{\ \ }}{+}}}}
\def\bfr{{\bf r}}
\def\bfw{{\bf w}}
\def\bfc{{\bf c}}
\def\hpsi{\hat \psi (\bfr)}
\def\hpsid{\hat \psi^\dagger (\bfr)}
\def\tpsi{\tilde \psi (\bfr)}
\def\tpsid{\tilde \psi^\dagger (\bfr)}

\draft
\title{Self-consistent effects of continuous wave output coupling of
atoms from a Bose-Einstein condensate}
\author{D. A. W. Hutchinson}
\address{
Clarendon Laboratory,
Department of Physics,
University of Oxford,
Parks Road,
Oxford OX1 3PU
}

\date{\today}

\maketitle

\begin{abstract}
We present a self-consistent mean field model of the extraction of
atoms from a Bose-Einstein condensate to form a CW atom laser. The
model is based upon the Hartree-Fock Bogoliubov equations within
the Popov approximation, modified by the inclusion of spatially
dependent source and sink terms, which lead 
to current flow within the condensate. The effects of this current
flow are investigated for traps containing Rubidium (repulsive 
effective interaction) and Lithium (attractive interaction) atoms.
The extra kinetic energy associated with this flow is shown to be
able to stabilise the condensate in the attractive case against
mechanical instability.
\end{abstract}

\pacs{PACS Numbers: 03.75.Fi, 05.30.Jp, 67.40.Db}

The possibility of producing an intense coherent beam of atoms from
a trap containing a sample of Bose-Einstein condensed atoms, a so
called atom laser, has attracted much interest in the time since
Bose-Einstein condensation was first achieved experimentally in a
dilute gas of atoms \cite{eric}. To our knowledge little or no work has appeared
however, investigating what effects extracting such a beam of atoms would
have on the properties of the condensate. In particular, does the 
extraction of a small proportion of the condensate over some reasonable
timescale drastically alter properties such as the condensate profile
and noncondensate fraction \cite{tony}? Given the great qualitative and quantitative
success of mean field theories \cite{strin}, specifically Hartree-Fock Bogoliubov (HFB)
theory \cite{hfb}, in determining the collective excitation spectrum, condensate
profile, noncondensate density and noncondensate fraction for trapped
Bose gases, we develop a version of the HFB formalism within the Popov
approximation \cite{popov} (which neglects two-body correlations) \cite{nick}, in which
we include a spatially dependent sink term, which extracts atoms from the
condensate. This corresponds to the output coupling of atoms from the
trap in some manner. We envisage some form of continuous Raman output
coupling \cite{mark}, but the specific mechanism is unimportant in this treatment. Also
included is a stimulated pumping term which corresponds to the Bose enhanced
scattering of atoms from the non-condensate into the condensate. The total
number of atoms in the trap is kept constant and the noncondensate and
condensate profiles are determined self-consistently. This corresponds to
an assumption that the trapped atoms are replenished at the same rate as
atoms are removed from the condensate via a thermal bath of atoms at a 
given temperature and that these 'new' thermal atoms equilibrate on a
timescale that is small compared to the characteristic timescale of the
trap.

This leads to a modified time-dependent Gross-Pitaevskii equation now
given by,
\begin{equation}
i \hbar \frac{\partial \Psi (\bfr , t)}{\partial t} = \left\{- \frac{\hbar^2}
{2M}\nabla^2 + V_{ext}(\bfr) + g[n_c(\bfr) + 2 \tilde n(\bfr)] -\frac{i \hbar}
{2} \gamma_c (\bfr) + \frac{i \hbar}{2} \Gamma \tilde n (\bfr) \right\}
\Psi (\bfr , t).
\end{equation}
Here we have made the usual decomposition of the Bose field operator, $\hat{
\psi}(\bfr, t)$, into condensate and noncondensate parts, i.e., $\hat{\psi}
(\bfr ,t) = \Psi (\bfr ,t) + \tilde{\psi}(\bfr ,t)$. The terms involving the
interaction strength, $g=4 \pi \hbar^2 a/M$, arise from the use of a contact
interaction, $g \delta(\bfr)$, where $a$ is the scattering length measured
for binary scattering in vacuo. The condensate and noncondensate densities
are given by $n_c$ and $\tilde n$ respectively. The factor of 2 before the
$\tilde n(\bfr)$ comes from the equivalence of the direct and exchange energies
in the Bose case when one uses a contact interaction.
The term involving $\gamma_c(\bfr)$ is the
sink term due to the extraction of condensate atoms via the output coupling
into the atom laser beam, with a coupling strength determined by $\gamma_c$.
The Bose enhanced pumping of atoms from the noncondensate into the condensate
is given by the term involving $\Gamma$, which is a constant determined 
self-consistently such that the number of atoms scattered 
from noncondensate to condensate per unit time is equal to the number of
atoms extracted from the condensate. Note that this form of the coupling
automatically includes the Bose enhancement factor.

We now look for a steady state solution for the self-consistent condensate
and non-condensate profiles by looking for solutions of the form 
$\Psi (\bfr ,t)=\Phi (\bfr) e^{-i\frac{\mu}{\hbar} t} e^{i \theta (\bfr)}$. Here
$\mu$ is the chemical potential and $\theta (\bfr)$ is a position dependent
phase. Normally the phase is an arbitrary constant and is set to zero without
loss of generality. The phase only becomes a relevant quantity in the context
of phase {\it differences} or gradients. In this problem, with spatially
dependent sink and source terms for condensate atoms, one does have a phase
gradient which corresponds to a flow of atoms within the condensate.

Making this substitution in the time-dependent Gross-Pitaevskii equation and
separating into real and imaginary parts, one obtains two equations. The first
is a modified time-independent Gross-Pitaevskii equation;
\begin{equation} 
\left\{- \frac{\hbar^2}
{2M}\nabla^2 + \frac{\hbar^2}{2M} \left ( \frac{\partial \theta (r)}{\partial r}
\right )^2 + V_{ext}(r) + g[n_c(r) + 2 \tilde n(r)] \right\} \Phi (r) = 
\mu \Phi (r),
\end{equation}
the second being the differential equation determining the phase gradient
\begin{equation}
\Phi (r) \frac {\partial^2 \theta (r)}{\partial r^2} + 2 \frac {\partial \Phi
(r)}{\partial r} \frac {\partial \theta (r)}{\partial r} + 2 \frac{\Phi (r)}
{r} \frac {\partial \theta (r)} {\partial r} + \frac {\hbar}{2} \left (
\Gamma \tilde n (r) - \gamma_c (r) \right ) = 0.
\end{equation}
Here we have made the additional simplifying assumption that the trap is
isotropic.

Similar analysis leads to the usual set of coupled HFB equations (within the
Popov approximation) given by
\begin{eqnarray}
\hat {\cal L} u_i (r) &-& g n_c (r) v_i (r) = E_i u_i (r) \\
\hat {\cal L} v_i (r) &-& g n_c (r) u_i (r) = -E_i v_i (r),
\end{eqnarray}
where the Hermitian operator, $\hat {\cal L}$, 
is modified by the addition of a kinetic energy term associated with the flow
of condensate atoms and is given by
\begin{equation}
\hat {\cal L} \equiv - \frac {\hbar ^2}{2 M} \nabla^2 + \frac {\hbar^2}{2M}
\left ( \frac {\partial \theta (r)}{\partial r} \right )^2 + V_{ext} (r) + 2g
(n_c (r) + \tilde n (r)) - \mu.
\end{equation}
The HFB equations define the quasiparticle excitation energies $E_i$ and
amplitudes $u_i$ and $v_i$. The condensate density is defined such that,
$n_c \equiv |\Phi (r)|^2$
and noncondensate density such that, 
\begin{equation}
\tilde n (r) \equiv \langle \tpsid \tpsi \rangle = \sum_i \left \{ |v_i (r)|^2
+ \left [ |u_i (r)|^2 + |v_i (r)|^2 \right ] N(E_i)
\right \},
\end{equation}
where $N(E_i)$ is the usual Bose factor with a temperature determined by that
of the external bath of atoms.

These now form a closed set of equations which can be solved self-consistently
giving the condensate and noncondensate densities, excitation spectrum and
phase gradient.

We now present results for two different regimes. Firstly we consider an
isotropic trap with $V_{ext} (\bfr) \equiv V_{ext} (r) = \frac {1}{2} M
\omega_0^2 r^2$, where $M(^{87}Rb)=1.44 \times 10^{-25}$ kg, and the trap
frequency $\nu_0 = \omega_0/2 \pi =200$ Hz. The scattering length for 
$^{87}Rb$ is positive (repulsive effective interatomic interaction) and
is given by $a \simeq 110 a_0 =5.82 \times 10^{-9}$ m. We consider a trap
containing 4000 atoms in the steady state at a temperature of 90 nK. Solving
self-consistently for the noncondensate fraction with no output coupling yields
2263 atoms in the condensate and 1737 atoms in the noncondensate at
this temperature, i.e. just less than 50 \% thermal depletion. We then extract
condensate atoms in a localised region at the centre of the trap via a
coupling $\gamma_c (r) = Ae^{-\sigma r^2}$. The total number of atoms extracted
in a characteristic time $\tau=1/\omega_0$ is given by,
\begin{equation}
N_{out}=\int d^3r
n_c(r) \left [ 1-e^{-\gamma_c \tau} \right ].
\end{equation}
We choose $\sigma = 2$ such 
that the full-width half-maximum (FWHM) of the coupling is of the order of
$1/3$ of the FWHM of the condensate. This enables one to induce moderately
large phase gradients (and hence current densities, $J=n_c \partial \theta
/ \partial r$) within the condensate. If one is considering a Raman output
coupling scheme, the FWHM would correspond to the focusing of the laser beams.
This is of course limited ultimately by the wavelength of the light used. With
the parameters used here the focusing would have to be on the order of 1 
$\mu$m, which is on the extreme limit of achievability. We would like to 
emphasise that we are looking at a trap containing only 4000 atoms (to which
we are limited numerically). Realistic traps are much larger now \cite{carl} and hence the
focusing in experiments can be much weaker and attain similar results 
qualitatively to those presented here.

In Fig 1 we present results for the condensate and noncondensate densities
and the current density for a range of amplitudes, $A$, of the coupling
strength. Note here that the current density is defined such that current
flow towards the centre of the trap is positive. For relatively weak
coupling the phase gradient established is small and the density profiles
of the condensate and noncondensate are essentially identical to those obtained
without any laser output. With increasing coupling strength, the current
densities grow monotonically (left panels) with a corresponding increase
in the effects upon the densities \cite{note}. The condensate becomes slightly depleted
at the centre and the noncondensate density increases slightly. This can
be thought of as an effective heating caused by the extraction of the laser
beam, but even in this extreme case where we have tried to maximise  the 
phase gradients by taking a small extraction region, the total increase
in the number of atoms in the noncondensate as opposed to the condensate
is from 1737 atoms with no extraction to 1822 atoms at the maximum
extraction rate when $N_{out} \simeq 100$ atoms in time $\tau$.

When the coupling strength is increased further (right panels) the effects
upon the densities become more pronounced and the condensate density at the
centre of the trap becomes depleted significantly. This actually leads to
a reduction in the rate of output of atoms from the trap as the coupling
strength is increased. This can also be seen from the fact that the current
density for the case with strongest coupling (dashed lines), has actually
decreased. There is therefore an optimal strength with which to couple
atoms from the condensate so that one can maintain a reasonable yield of
atoms without depleting the condensate density excessively. This is equivalent
to saying that there is a maximum flow rate for atoms within the condensate.
 
The second case we wish to consider is the case when the trapped atoms have
a negative scattering length, i.e. attractive effective interactions. We
therefore consider a trap with $\nu_0 = \omega_0/2\pi=150$ Hz and $M(^7Li)
= 1.16 \times 10^{-26}$ kg. The s-wave scattering length for $^7Li$ is given
by $a=-1.44 \times 10^{-9}$ m. The parameters here correspond closely to
those of the Rice group \cite{randy}.

It is well known that condensates with attractive interactions are only 
metastable at best and above some critical condensate number become unstable
to mechanical collapse, which is characterised by the $l=0$ collective mode
going soft, i.e. the lowest $l=0$ excitation frequency goes to zero \cite{stoof}. The
critical number for the above parameters has been shown to be $N_c=1241$
atoms at $T=0$. It has also been shown that increasing the temperature
decreases the stability and decreases the critical number $N_c$. This decrease
manifests itself in a decrease of the $l=0$ mode frequency with temperature.
For example, with 1150 atoms in the condensate and a total number of atoms in
the trap determined self-consistently from the HFB equations, the $l=0$ mode 
frequency drops from $\omega_{l=0}=1.523 \omega_0$ at $T=0$ to $\omega_{l=0}
=1.428 \omega_0$ at $T=75$ nK, with 1312 atoms in the noncondensate, viz.
a total of 2462 atoms in the trap.

The (meta) stability of the condensate is due to the kinetic energy associated
with the trapping potential - there is no stable condensate in the uniform case
with attractive interactions. The extraction of atoms via an atom laser sets
up a current flow within the condensate which has the effect of increasing the
kinetic energy. It is therefore feasible that the extraction of atoms via some
method of localised output coupling could stabilise the condensate against
mechanical collapse. We therefore apply our model to the case with 1150 atoms
in the condensate at 75 nK. In this case $\sigma=256$ which corresponds to a
FWHM of 0.4 $\mu$m. This puts one into the uv region for Raman output coupling,
which is infeasible. However for this trap a very small output region is 
required as the characteristic size of the condensate is small. To 
experimentally investigate these effects one would need to use a trap with a
smaller confining potential and hence larger (less dense) condensate. The
amplitude of the coupling is 2500 $\omega_0$ which yields a cw laser output
of 129.46 atoms/$\tau$. The condensate, noncondensate (magnified by a factor of
20) and current densities are shown in Fig 2. The peak in the current density
approximately corresponds, in terms of position to that of that of the point of
inflection in $\gamma_c (r)$. The relative energy scales of the kinetic energy
due to the current flow and the scale of the interaction energy in this case
is of the order
\begin{equation}
\frac {\hbar^2}{2M} \left ( 
\frac {\partial \theta }{\partial r}   \right
)^2_{max}/ g n_c^{max} \simeq 0.2.
\end{equation}
Even with this relatively small addition to the 
kinetic energy the frequency of the $l=0$ collective excitation has been 
increased form $\omega_{l=0}= 1.428 \omega_0$ to $\omega_{l=0}=1.464 \omega_0$.
This represents a significant decrease in the softening of the mode and a
strong indication of the feasibility of stabilising the condensate against
collapse. 

In conclusion, we have developed a model based upon well tried self-consistent
mean field theory that incorporates the effects of continuous extraction of 
atoms from the condensate and pumping via scattering of atoms from the 
noncondensate to condensate. We have investigated the effects of the induced
current within the condensate upon the steady state properties of both the
condensate and noncondensate. For trapped atoms with repulsive interactions
and moderate extraction rates, the effects were not found to be excessively
damaging and it appears likely that atoms could be extracted by means of some
form of output coupler in a continuous manner, without significantly effecting
the steady state properties of the condensate.

In the case of trapped atoms with attractive interactions the spatially
localised extraction of an output beam of condensate atoms, setting up a
flow of current within the condensate was found to increase the kinetic
energy of the condensate. This was found to reduce the softening of the
$l=0$ mode through thermal effects, hence stabilising the condensate
against mechanical collapse. It is hoped that for larger traps (weaker
confinement) larger phase gradients may be created which, further increasing
the kinetic energy, may be able to increase the critical number of
condensate atoms for given trap parameters beyond that of the $T=0$ case
with no extraction.

This work was performed at the University of Auckland during a Royal Society
Study Visit and we gratefully acknowledge the Royal Society (London),
University of Auckland Research Committee, the Marsden Fund of the Royal
Society of New Zealand and the UK EPSRC for funding. Thanks go to Tony Wong,
Dan Walls and Keith Burnett for many useful discussions and very special thanks
in general to the quantum optics group at the University of Auckland for their 
warm hospitality.

{

\begin{figure}
\caption{Condensate and noncondensate densities in units of $d^{-3}$ and
current densities ($\omega_0 d^{-2}$) for a trapped gas of 4000 $^{87}Rb$
atoms at $T=90$ nK with a spatially dependent output coupling rate given
by $\gamma_c = Ae^{-\sigma r^2}$. $\sigma=2$. Left panels; $A=10$ (solid
line), $A=80$ (dotted line), $A=120$ (dashed line). Right panels; $A=160$
(solid line), $A=200$ (dotted line), $A=240$ (dashed line).
}
\end{figure}

\begin{figure}
\caption{
Condensate, noncondensate and current densities for a trapped gas of $^7Li$
atoms at $T=75$ nK with 1150 atoms in the condensate and Gaussian output
coupling strength $\gamma_c (r)$ centred at the origin.
}
\end{figure}

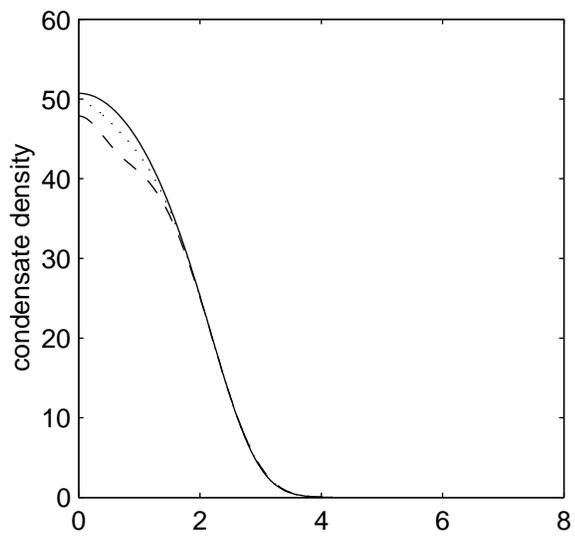
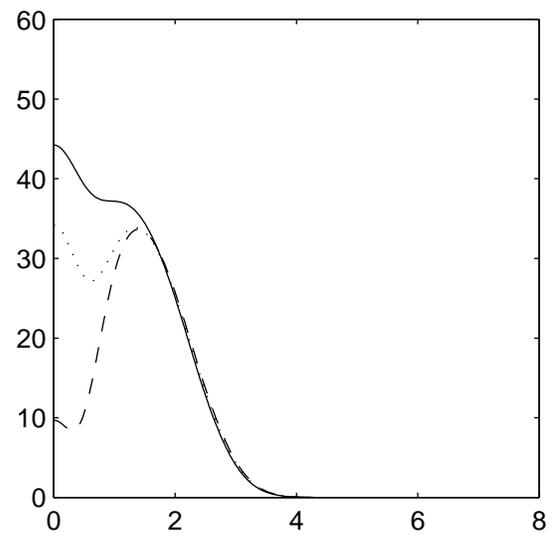
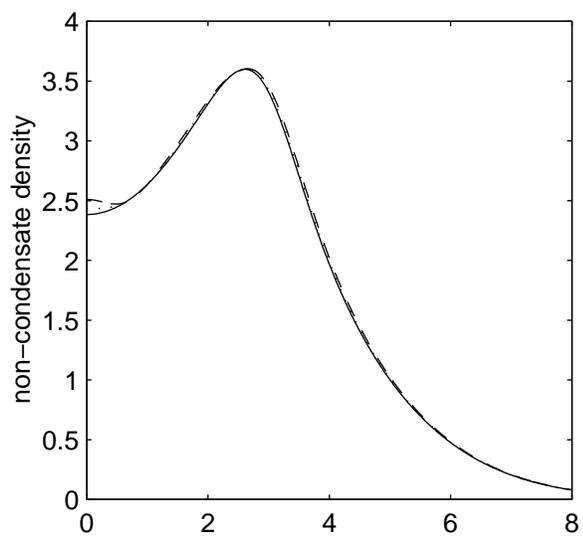
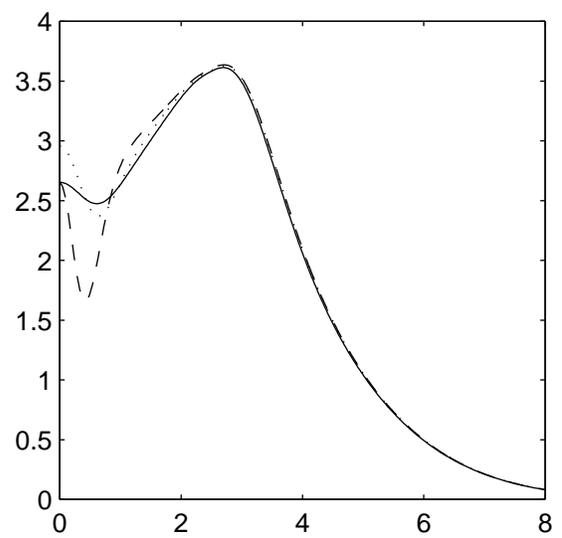
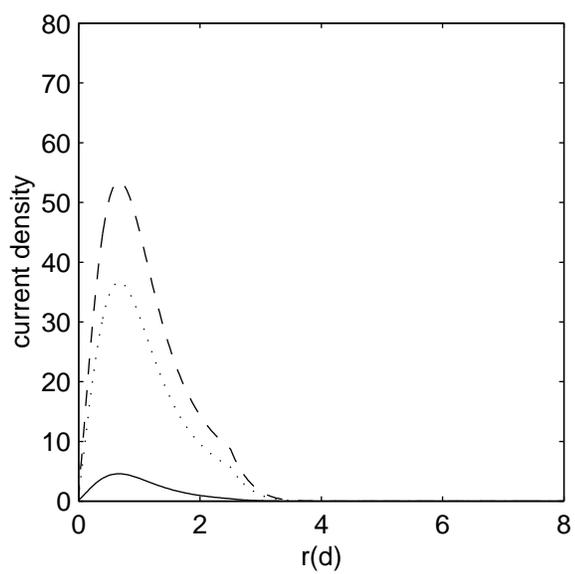
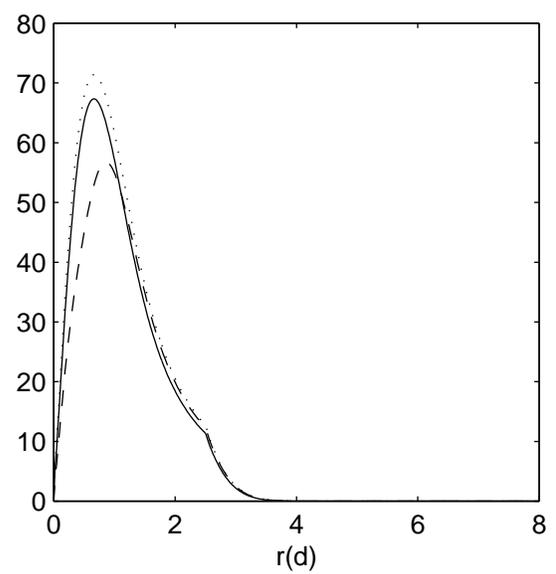
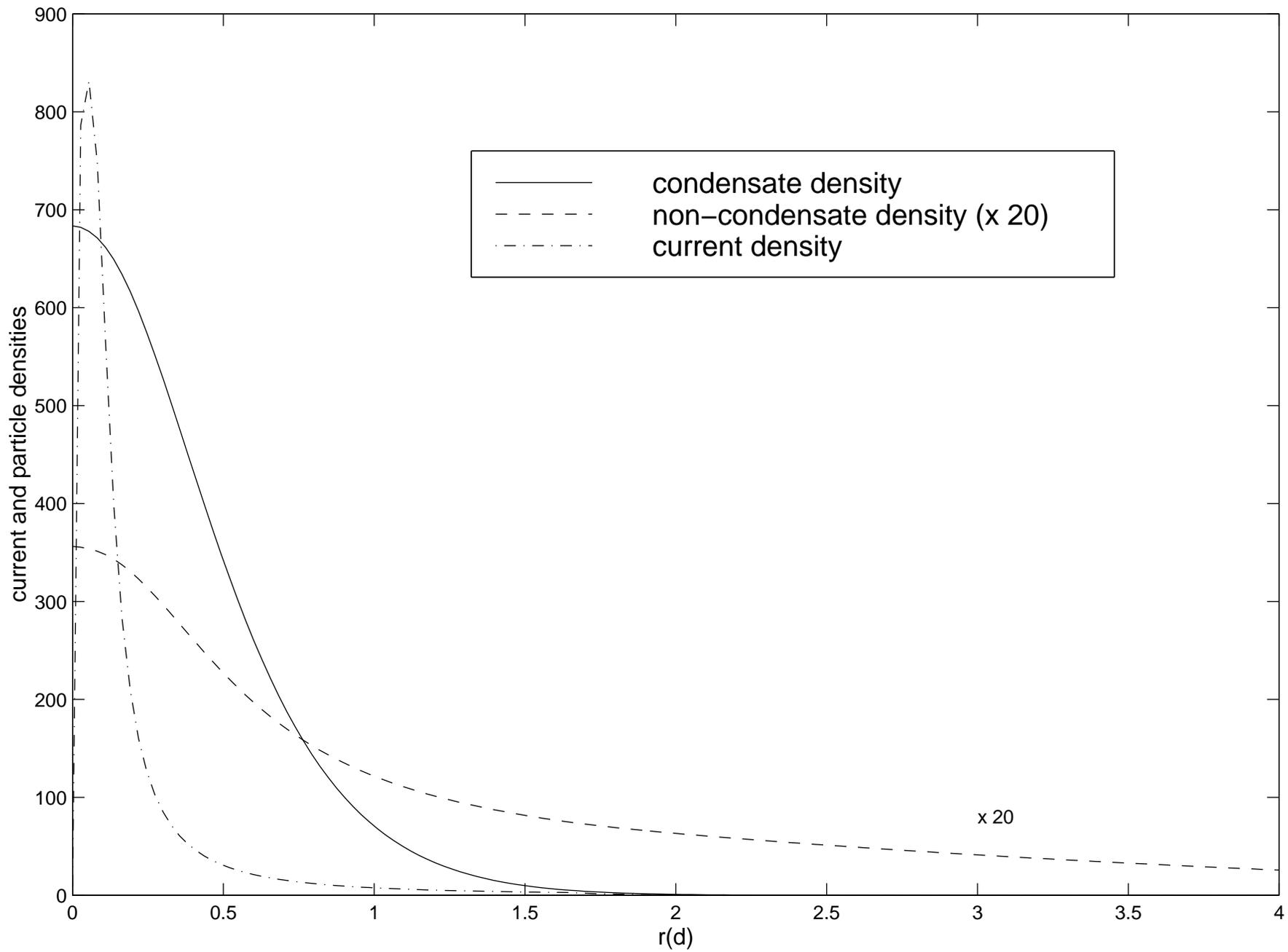
\end{document}